\documentstyle[11pt,epsfig]{article}
\begin{document}
\def\boxit#1{\vcenter{\hrule\hbox{\vrule\kern8pt
      \vbox{\kern8pt#1\kern8pt}\kern8pt\vrule}\hrule}}
\def\Boxed#1{\boxit{\hbox{$\displaystyle{#1}$}}} 
\def\sqr#1#2{{\vcenter{\vbox{\hrule height.#2pt
        \hbox{\vrule width.#2pt height#1pt \kern#1pt
          \vrule width.#2pt}
        \hrule height.#2pt}}}}
\def\square{\mathchoice\sqr34\sqr34\sqr{2.1}3\sqr{1.5}3}
\def\Square{\mathchoice\sqr67\sqr67\sqr{5.1}3\sqr{1.5}3}
\def\lambdabar{{\mathchar'26\mkern-9mu\lambda}}
\def\thrdotovervx{\buildrel\textstyle...\over v_x}
\def\thrdotovervy{\buildrel\textstyle...\over v_y}
\title{\bf Zitterbewegung in External Magnetic Field: Classic versus Quantum Approach}

\author{{\small Mehran Zahiri--Abyaneh}\footnote{mehran.z.abyaneh@gmail.com} \ {\small and\
         Mehrdad Farhoudi}\footnote{m-farhoudi@sbu.ac.ir}\\
        {\small Department of Physics, Shahid Beheshti University, G. C.,
        Evin, Tehran 19839, Iran}}
\date{\small February 25, 2011}
\maketitle
\begin{abstract}
We investigate variations of the Zitterbewegung frequency of
electron due to an external static and uniform magnetic field
employing the expectation value quantum approach, and compare our
results with the classical model of spinning particles. We
demonstrate that these two so far compatible approaches are~not in
agreement in the presence of an external uniform static magnetic
field, in which the classical approach breaks the usual symmetry
of free particles and antiparticles states, i.e. it leads to CP
violation. Hence, regarding the Zitterbewegung frequency of
electron, the classical approach in the presence of an external
magnetic field is unlikely to correctly describe the spin of
electron, while the quantum approach does, as expected. We also
show that the results obtained via the expectation value are in
close agreement with the quantum approach of the Heisenberg
picture derived in the literature. However, the method we use is
capable of being compared with the classical approach regarding
the spin aspects. The classical interpretation of spin produced by
the altered Zitterbewegung frequency, in the presence of an
external magnetic field, are discussed.
\end{abstract}
\bigskip
\section{Introduction}
\indent

As velocity and momentum in the Dirac theory~\cite{dirac2-1928}\
are related in a different manner than in non--relativistic
mechanics, it leads~\cite{Dirac58}--\cite{Merzbacher98} to the
fact that the average speed of spin--${1/2}$ particles must be
less than $c$, while the instantaneous speed is always $\pm c$ in
this theory. Indeed, by using an arbitrary state of a
spin--${1/2}$ free particle in which it is localized in space, one
can show~\cite{huang1952} that the expectation value of the
position, even in the absence of external field, includes a
time--dependent part which originates from an interference between
the positive and negative energy states. Hence, the motion of free
particle has a peculiar oscillatory component with a very high
frequency of $2mc^2/\hbar\simeq 1.6\times {10}^{21} {\rm s}^{-1}$\
for electrons with the rest mass m -- which secures $\pm c$ as the
eigenvalues of instantaneous speed. That is, a free particle obeys
a rapid oscillatory motion with the speed $c$ around the center of
mass while moving like a relativistic particle with velocity ${\bf
p}/m$. This angular frequency is referred to $\omega_{_{\rm
zbw}}\equiv 2mc^2/\hbar$ after Schr\"odinger who called this
motion \emph{Zitterbewegung}\ ({\bf
Zbw})~\cite{schrodinger30-31,barut-etal81a} -- trembling or
quivering motion -- which he theoretically contrived it after
Dirac had employed his equation for free relativistic electron in
vacuum. He interpreted this as a fluctuation in the position of
electron. The amplitude of the Zbw has also been indicated to be
of the order of the Compton wavelength,
$\lambdabar_c=\hbar/(mc)\simeq 3.86\times {10}^{-13} {\rm m}$ for
electrons. Now, the well--known phenomenon of the Zbw, caused by
the interference between the positive and negative energy
components of a wave packet, has been described in many textbooks
as well~\cite{Dirac58}--\cite{Merzbacher98}.

It is also well--known~\cite{Dirac58}--\cite{Merzbacher98} that as
the position operator in the Dirac theory is not a one--particle
observable, the path of an electron can only be indicated up to
the order of its Compton wavelength. Indeed, the best localized
wave packet exclusively consisting of plane wave components with
positive or negative energy has characteristic linear dimension of
about $\lambdabar_c$~\cite{newton-etal49}. However, the Zbw for
free electrons is not an observable motion and has never been
directly detected~\cite{rusin-etal2006} (nevertheless, see
discussion in Sec.~$3$). Actually, with certain initial
conditions, one can show~\cite{huang1952} that all components of
$<\!\mbox{\boldmath$r$}\!>$ vanish upon integration, which
indicates that the total contribution of all waves vanishes (see
Sec.~$2$). Hence, in these conditions, the center of the wave
packet as a whole remains at rest, as experimental evidence rules
out the possibility that electron is an extended
body\rlap.\footnote{Indeed, scattering experiments have revealed
that the size of electron is less than $10^{-18} {\rm
m}$~\cite{bender84}.}\
 Though, the consequence of the Zbw is that it is
impossible to localize electron better than to a certain finite
volume.

The above argument leaves open the possibilities that electron may
be regarded as a point charge which generates spin and magnetic
moment by an intrinsic local motion~\cite{hestenes93}. Actually,
in the literature, people have tried to analyze some picture about
the detailed motion of the Zbw and to interpret the structural
feature of electron. The motivation has mainly come from the close
connection between the Zbw and the intrinsic magnetic moment of
electron. That is, using the current density expression of a free
electron, one can
infer~\cite{gordon27frenkel341,gordon27frenkel342} that the
classical expression for the magnetic moment has its root in the
Zbw (see also Sec.~$2$). Hence, a temptation that the magnetic
moment and the spin of electron are generated by some sort of the
local circular motion of electron, i.e. mass and charge,
\cite{huang1952,hestenes93,barut-etal81b,hestenes90,hestenes2009}.

Indeed, many classical treatments of a free Dirac electron have
been formed based on this picture for the Zbw, see
Refs.~\cite{MHWB1}--\cite{SalesiRecami2000} and references
therein. A significant achievement toward understanding the Zbw
has been made~\cite{PFetalT1}--\cite{PFetalT3} when it was shown
that in the absence of external fields the Dirac Hamiltonian can
be transformed into a form in which the positive and negative
electron energies are decoupled. In this new representation,
electron is not a point--like particle and has the size of about
$\lambdabar_c$. The interpretation of these two pictures has been
considered in
Refs.~\cite{huang1952,barut-etal81a,newton-etal49},~\cite{FetalBetalL1}--\cite{Krekora-etal2004},
and the compatibility of the Zbw interpretation with the details
of the Dirac theory has also been
demonstrated~\cite{hestenes90,hestenes2009,BarutZanghi84}.

Huang~\cite{huang1952} obtained this picture from the Dirac
equation and claimed that it is not an unreasonable one. Actually,
using an explicit form for the wave packet with specified initial
condition to construct the wave packet mainly from positive or
negative waves, he showed that the contribution of each plane wave
to the motion of the wave packet is an orbit whose projection on
the plane perpendicular to the direction of the spin is a circle
of radius $\lambdabar_c/2$. Then, using the method of the Gordon
decomposition~\cite{gordon27frenkel341,gordon27frenkel342}, he
also justified that the intrinsic spin of electron may be
considered as an ``orbital angular momentum'' due to the Zbw. That
is, the current produced by the Zbw actually causes the intrinsic
magnetic moment of the particle\rlap,\footnote{The same issue, but
in the presence of an external magnetic field, is studied at the
end of Sec.~$2$.}\
 and hence, the total
magnetic moment of electron is represented by both the orbital and
the intrinsic angular momenta with the correct gyromagnetic $g$
factor.

A classical analogous model for spinning electron, which has been
claimed to be a natural classical limit of the Dirac
theory~\cite{SalesiRecami2000}, has been
presented~\cite{BarutZanghi84} by Barut and Zanghi ({\bf BZ}) in a
particular five dimensional space--time (which we refer to as BZ
approach). They introduced a classical Lagrangian (see
relation~(\ref{Lag})) in a manifestly covariant form using a pair
of spinorial variables (representing internal degrees of freedom)
besides the usual pair of dynamical variables $(x^{\mu},
p^{\mu})$, as functions of the proper time parameter measured in
the center of mass frame (see
Refs.~\cite{SalesiRecami2000,RecamiSalesi98} and references
therein). The general solution of the motion for free electrons
includes an oscillatory part as well, which denotes the presence
of the Zbw for the spinning particles (see Sec.~$3$). That is, the
Zbw occurs because the motion of particle does~not coincide with
the motion of the center of mass~\cite{Salesi2002}, and a spinning
particle appears as extended object~\cite{RecamiSalesi98}. It has
also been found that free polarized particles have circular
motions in the plane perpendicular to the direction of the spin,
with the Zbw frequency and half the Compton wavelength radius in
the center of mass frame.

After quantization, the BZ approach has also been indicated to
describe the Dirac electron~\cite{Barut-pavsic}. In fact, this has
been originated from the classical Hamiltonian corresponding to a
classical Lagrangian devised by BZ~\cite{BarutZanghi84} (here
shown with relation~(\ref{Lag})). Then, it has been converted to a
quantum version with reinterpretation of the spinorial variables
as Dirac field spinors~\cite{RecamiSalesi1995,Pavsic-elal1998}.
Also, by recasting the BZ approach into the language of Clifford
formalism\rlap,\footnote{In which they have also claimed that this
algebra allows various independent classical approaches of the
spin to be unified.}\
 quantization of the theory has been
accomplished~\cite{Pavsic-elal1998}.

The significant correspondence between the Dirac theory and BZ
approach is intriguing and invites one to discover up to where
this compatibility lasts, and actually whether there is~not any
conflict between them. To pursue such a query, we purpose to study
the particle's Zbw in the presence of an external uniform static
magnetic field. Our stimulated hope in exploring this procedure is
that theoretical understanding of the Zbw may shed light on the
nature of spin of elementary particles.

In the standard textbooks~\cite{Dirac58}--\cite{Merzbacher98}, the
Zbw has only been examined in details for free spinning particles
without considering any interactions. Though, it has been
indicated that electron in the hydrogen atom, i.e. in the presence
of the Coulomb potential, also displays the Zbw which gives rise
to the so called Darwin term~\cite{Dirac58}--\cite{Merzbacher98}.
Interestingly enough, the phenomenon of Zbw for electrons has been
shown~\cite{rusin-etal2006},~\cite{ZCetalFetalZJetalZZSetalK1}--\cite{BrusheimXu2008}
to occur in non--relativistic case and in solids in the absence of
external fields. Even, an analogous quantum--relativistic effect
in a single trapped ion has been
studied~\cite{LamataLeonSchatzSolano}. In these conditions, it has
much lower angular frequency and much larger amplitude, and it may
lead easier to experimental detection (also see Sec.~$3$). Also,
in
Refs.~\cite{SalesiRecami2000,BarutZanghi84,Villavicencio-etal2000}
and intuitively in Ref.~\cite{holten92,holten93}, the phenomenon
of Zbw for electrons in the presence of an external magnetic field
has been established. Hence, the Zbw is~not a special
characteristic of plane wave solution, and does occur whenever
positive and negative energy states are
mixed~\cite{hestenes90,Villavicencio-etal2000}.

Actually, by interpreting the interference between positive and
negative energy states, the Zbw may be considered to arise in
electron--positron pair creation and
annihilation~\cite{thirring58} -- which manifests the Zbw as a
purely relativistic phenomenon. On the other hand, the Zbw is
considered to be the origin of the spin only in the
non--relativistic domain. By reformulating the Dirac theory, it
has been claimed~\cite{hestenes90,hestenes2009}\footnote{Also, see
Ref.~\cite{Pavsic-elal1998}.}\
 that the Zbw is a
phenomenon which manifests itself in every application of quantum
mechanics, i.e. either in the non--relativistic domain and/or in
the presence of an arbitrary electromagnetic interaction. Indeed,
it has been claimed~\cite{hestenes93,hestenes90} that the most
features of quantum mechanics are manifestations of the Zbw, e.g.
the complex phase factor in the electron wave function as physical
representation of rotation can be associated directly with the
Zbw. However, we should mention the
assertion~\cite{Krekora-etal2004} that quantum field theory
prohibits the occurrence of the Zbw for an electron, and in
principle, there is no difficulty to localize an electron narrower
than the Compton wavelength if it is involved with a positron.

Regarding the question of an electron in vacuum in the presence of
an external magnetic field, the references~\cite{SalesiRecami2000}
and~\cite{Villavicencio-etal2000} are of much interest to us. This
is in fact the topic that we purpose to deal with by making
recourse to the Huang quantum approach~\cite{huang1952} where the
next section is devoted to. There, we contrast the results with
those two investigations, particularly with the more accurate
solutions that we will indicate by using
Ref.~\cite{SalesiRecami2000} in where the BZ classical approach
has been employed to study the issue. Actually, in Sec.~$2$, the
expectation value approach has been utilized to find changes of
$\omega_{_{\rm zbw}}$ in an external uniform static magnetic
field. In Sec.~$3$, we review the classical approach of
Ref.~\cite{SalesiRecami2000} and infer some improvements. It will
be shown that the results derived via the expectation value
quantum approach are different from the corresponding ones
obtained via the classical approach from
Ref.~\cite{SalesiRecami2000}. Finally, a brief discussion and
conclusion is given in the last section.

\section{Frequency Shift via Expectation Value Approach}
\indent

Huang~\cite{huang1952} used solutions to the Dirac equation and
the expectation value of velocity to study the issue of free
electron's Zbw. In this work, we evaluate the electron's Zbw in
the presence of an external uniform static magnetic field
utilizing his method. Thus, we first very briefly review some
results for the free particle and then, we will extend the method
to our case.

To examine the average position of an electron, which is localized
in space, a general Dirac wave packet, expanded in terms of
momentum eigenfunctions, as
\begin{equation}\label{Wav}
\!\!\!\!\!\Psi({\mbox{\boldmath$r$}},t)=h^{-3/2}\int{\Bigl[C_+({\mathbf{p}})\exp(-i\omega
t) +C_{-}({\mathbf{p}})\exp(i\omega
t)\Bigr]\exp\Bigl(i\frac{{\mathbf{p}}\cdot{\mbox{\boldmath$r$}}}{\hbar}\Bigr)
d^{3}p}
\end{equation}
can be used. This wave packet includes both negative and positive
energies, and $C_{+}({\mathbf{p}})$\
$\left(C_{-}({\mathbf{p}})\right)$ is a linear combination of the
spin--up and spin--down amplitudes of the free particle Dirac
waves with momentum $\textbf{p}$ and positive (negative) energy.
Using this wave packet and calculating
$<\!\dot{\mbox{\boldmath$r$}}\!>$, and consequently
$<\!\mbox{\boldmath$r$}\!>$ with a suitable initial condition
having a well--defined initial spin direction, it has been
shown~\cite{huang1952} that each plane wave in the above Fourier
decomposition contributes a circular motion in the plane
perpendicular to the direction of the electron spin with radius
$\lambdabar_c/2$ and frequency $\omega_{_{\rm zbw}}$. This leads
to the internal spin and hence, the magnetic moment of the Dirac
particle.

Now, the Dirac equation for an electron in an external
electromagnetic field is
\begin{equation}\label{Ham}
i\hbar\frac{\partial\Psi}{\partial
t}=\left[c\alpha_i(p^i-\frac{e}{c}A^i) +\beta mc^2+eA_0\right]\Psi
,
\end{equation}
where the matrices $\mbox{\boldmath$\alpha$}$ and $\beta$ are
defined as\footnote{In terms of the Dirac (gamma) matrices,
$\gamma_a$, they are $\beta\equiv\gamma^0$ and
$\alpha_i\equiv\gamma^0\gamma_i$ for $i=1, 2, 3$.}
\begin{eqnarray}\label{Pau}
\alpha_i=\left(
                 \begin{array}{cc}
                   0 & \sigma_i \\
                  \sigma_i & 0 \\
                 \end{array}
               \right)\
               \qquad\ \textrm{and}\qquad\ \beta=\left(
                 \begin{array}{cc}
                   I & 0 \\
                   0 & -I \\
                 \end{array}
               \right),
\end{eqnarray}
with $\sigma_i$ and $I$ as the $2\times2$ Pauli matrices and the
unit matrix, respectively. We proceed with special case of an
external uniform static magnetic field without a scalar Coulomb
field (i.e. $A_0=0$). Then, using the standard abbreviation
$\pi_i\equiv p_i-eA_i/c$ in the Gaussian units -- i.e. defining
the gauge--invariant momentum operator correspondence
$\pi_\mu\rightarrow -i\hbar D_{_{\mu}}$ in terms of the covariant
derivative\footnote{In a general case, the rate of change of this
generalized operator, via the Heisenberg picture, gives the
Lorentz force, i.e. $\dot{\mbox{\boldmath$\pi$}}=i[H,
{\mbox{\boldmath$\pi$}}]/\hbar-(e/c)\partial {\mathbf{A}}/\partial
t=e({\bf E}+{\mbox{\boldmath$\alpha$}}\times{\bf B})$.}\
 -- the Dirac equation (\ref{Ham}) reads $i\hbar\partial\Psi/\partial
t=\left(c\,\alpha _i\pi^i +\beta mc^2\right)\Psi$. Making a
general wave solution {\it ansatz\/}\footnote{We should emphasis
that due to the presence of an external magnetic field the wave
packet $\Psi$ is~not an eigenfunction of neither the operator
$\pi_i$ nor the operator $p_i$.}
\begin{equation}\label{Anz0}
\Psi(\mbox{\boldmath$r$}, t)=\left(
  \begin{array}{c}
    \phi(\mbox{\boldmath$r$}) \\
  \chi(\mbox{\boldmath$r$})\\
  \end{array}
\right)\exp -i\left(\frac{E_{_{\rm B}} t}{\hbar}\right),
\end{equation}
with $E_{_{\rm B}}$ as the particle energy in the presence of an
external magnetic field, hence the Dirac equation yields
\begin{eqnarray}\label{Equ}
\left(E_{_{\rm B}}+mc^2\right)\chi=c\sigma_i \pi^i\phi\qquad\
\textrm{and}\qquad\
 \left(E_{_{\rm B}}-mc^2\right)\phi=c\sigma_i\pi^i\chi\, .
\end{eqnarray}

A uniform magnetic field is derivable from a vector potential $
{\mathbf{A}}=({\mathbf{B}}\times{\mbox{\boldmath$r$}})/2$, which
for $\mathbf{B}$ along the $z$--direction, becomes $
{\mathbf{A}}=-(yB\hat{x}-xB\hat{y})/2$. Such a field, without loss
of generality, can also be achieved by choosing a vector potential
as $A_x=0=A_z$ and $A_y=xB$. This way, one can make {\it
ansatz}~\cite{ItzyksonZuber}
\begin{equation}\label{Anz}
\phi(\mbox{\boldmath$r$})=h(x)\exp [i(p_y y+p_z z)/\hbar]
\end{equation}
to eliminate $\chi$ in equations~(\ref{Equ}) and then finds that
$h(\xi)\propto H_n(\xi)\exp( {-\xi^2/2})$, where $H_n$ represents
the Hermit polynomials. Noting that positive energy corresponds to
electron with negative charge ($-|e|$) while negative energy
describes positron with positive charge ($|e|$), and taking the
sign of $B$ such that $-|e|B>0$, one has
$\xi=\sqrt{-|e|B/c}\left(x\pm cp_y/|e|B\right)$ with the plus
(minus) sign for positive (negative) energy.

Thus, the energy spectrum of a particle with a magnetic moment
\mbox{\boldmath$\mu$} in a strong external magnetic field along
the $z$--direction, employing the cylindrical coordinates and
neglecting back--reaction of the motion of charge on the field,
takes the form~\cite{ItzyksonZuber,holten92}
\begin{equation}\label{Eng}
E_{_{\rm ln}}^2=m^2c^4+p_z^2c^2+ceB(n-l+1-2s_{z}),
\end{equation}
where $p_z=\hbar k$, $s_{z}=\pm\hbar/2$, $n$ is a non--negative
integer and $L_z=\hbar l$, while $l$ is restricted to values
$l=-n, -n+2, \cdots, n-2, n$.

The term $n-l+1$ represents contribution from the orbital part of
the magnetic moment, whereas we purpose to deal with effects of an
external magnetic field only on the internal spin. Thus, from the
spectrum~(\ref{Eng}), one can infer~\cite{holten93}
\begin{equation}\label{Eng1}
E_{_{\rm B}}^2=m^2c^4+p_z^2c^2-2ceBs_{z}\, .
\end{equation}
Defining $\sigma_{\pm}=\pm1$ as the eigenvalues of the operator
$\sigma_z$ with positive (negative) sign representing the up
(down) spin state, and reminding that positive (negative) energy
relates to electron (positron), relation~(\ref{Eng1}), in the weak
field approximation and non--relativistic limit (justified below
Eq.~(\ref{vawf})), reads
\begin{equation}\label{Omg}
\frac{E_{_{\rm B\pm}}}{\hbar}\simeq \pm\left(\frac{
mc^2}{\hbar}\pm\frac{\mid\! e\!\mid B \sigma_{\pm}}{2m
c}\right)\equiv \pm\left(\omega\pm\Omega\sigma_{\pm}\right),
\end{equation}
where $\omega\equiv\omega{_{\rm zbw}}/2$,\ $\Omega\equiv\,\mid\!
e\!\mid\!  B/2m c=\omega_c/2$,\ $\omega_c$ is the usual classical
cyclotron frequency. The weak field approximation means
$\Omega\ll\omega$ or $\omega_c\ll\omega_{_{\rm zbw}}$, which is a
reasonable approximation even with the highest magnetic field
achievable in the current laboratories. One should also note that,
having the external magnetic field along the $z$--direction with
the proposed vector potential, the only altered generalized
momentum is $\pi_y$, while $\pi_z=p_z$ and  $\pi_x=p_x$. In
addition, in order to justify use of Eq.~(\ref{Eng}) and also
simplify later on calculations, we ignore the non--commutativity
of $\pi_x$ and $\pi_y$ (justified below Eq.~(\ref{commut})).

As we are interested to find effects of an external magnetic field
on the $\omega_{_{\rm zbw}}$, following Huang's approach, it is
more appropriate and instructive to solve these equations by the
initial spinoral eigenstates. That is, using
$\phi^\uparrow=(h^{\uparrow}_+, 0)$ and $\phi^\downarrow=(0,
h^{\downarrow}_+)$ as the initial positive energy eigenstates for
the spin--up and spin--down, and $\chi^\uparrow=(h^{\uparrow}_-,
0)$ and $\chi^\downarrow=(0, h^{\downarrow}_-)$ as the initial
negative energy eigenstates for the spin--up and spin--down,
respectively. Then, by relation~(\ref{Omg}) and {\it
ansatz}~(\ref{Anz}), a general solution of Eqs.~(\ref{Equ})
becomes
\begin{eqnarray}\label{Ans}
\Psi({\mbox{\boldmath$r$}},t)=h^{-3/2}\int \Biggl\{{\cal A}\left(
              \begin{array}{c}
                \phi^\uparrow \\
              k_{+}^\uparrow\mbox{\boldmath$\sigma$}\cdot\mbox{\boldmath$\pi$}\phi^\uparrow  \\
              \end{array}
            \right)\exp\Bigl[-i(\omega+\Omega)t\Bigr]
            +{\cal B}\left(
                \begin{array}{c}
                 \phi^\downarrow \\
              k_{+}^\downarrow\mbox{\boldmath$\sigma$}\cdot\mbox{\boldmath$\pi$}\phi^\downarrow \\
                \end{array}
              \right)\exp\Bigl[-i(\omega-\Omega)t\Bigr]\cr
              +\,{\cal C}\left(
                  \begin{array}{c}
                     \!\!k_{-}^\uparrow\mbox{\boldmath$\sigma$}\cdot\mbox{\boldmath$\pi$}\chi^\uparrow\!\! \\
              \chi^\uparrow \\
                  \end{array}
                \right)\exp\Bigl[i(\omega-\Omega)t\Bigr]
            +{\cal D}\left(
                \begin{array}{c}
                  \!\!k_{-}^\downarrow\mbox{\boldmath$\sigma$}\cdot\mbox{\boldmath$\pi$}\chi^\downarrow\!\! \\
                \chi^\downarrow\\
                \end{array}
              \right)\exp\Bigl[i(\omega+\Omega)t\Bigr]\,\Biggr\}\exp [i(p_y y+p_z z)/\hbar] \,d^3\pi,
\end{eqnarray}
where ${\cal A}$, ${\cal B}$, ${\cal C}$ and ${\cal D}$ are
complex functions of $\mbox{\boldmath$\pi$}$ and $k$'s. The
notation $k_{\pm}$ for positive and negative energies in the
absence of external fields are $k_{\pm}=\pm c/(E+mc^2)$ with
$E=(p^2c^2+ m^2c^4)^{1/2}$~\cite{huang1952}, while in our
situation, with the aid of relation~(\ref{Omg}), it reads
\begin{equation}\label{Norm1}
 k_{-}^\uparrow= -k_{+}^\downarrow=\frac{-
 c}{2mc^2-\hbar\Omega}\equiv- K_{1}
 \qquad\quad {\rm and}\qquad\quad
 k_{-}^\downarrow=-k_{+}^\uparrow =\frac{- c}{2mc^2+\hbar\Omega}\equiv-K_{2}\, .
\end{equation}

More explicitly, comparing solution~(\ref{Ans}) with the one in
the absence of magnetic field -- namely the plane wave
packet~(\ref{Wav}) -- reveals that one has
\begin{eqnarray}\label{UJ1}
\cases{
       \exp(i\omega{t})\rightarrow{\exp[i(\omega-\Omega)t]}\ &\textrm{for negative energy}\cr
       \cr
       \exp(-i\omega{t})\rightarrow{\exp[-i(\omega+\Omega)t]}\quad &\textrm{for positive energy}\cr}
\end{eqnarray}
for the spin--up states, and
\begin{eqnarray}\label{DJ1}
\cases{
       \exp(i\omega{t})\rightarrow{\exp[i(\omega+\Omega)t]}\ &\textrm{for negative energy}\cr
       \cr
       \exp(-i\omega{t})\rightarrow{\exp[-i(\omega-\Omega)t]}\quad &\textrm{for positive energy}\cr}
\end{eqnarray}
for the spin--down states. Let us write the wave
packet~(\ref{Ans}) in the form
\begin{eqnarray}\label{Ansc}
\Psi({\mbox{\boldmath$r$}},t)\!\!\!\!&=&\!\!\!h^{-3/2}\int\Biggl\{
C_{+}^\uparrow(\mbox{\boldmath$\pi$})h^\uparrow_+\exp\Bigl[ -i
(\omega+\Omega) t\Bigr]+
C_{+}^\downarrow(\mbox{\boldmath$\pi$})h^\downarrow_+\exp\Bigl[ -i
(\omega-\Omega) t\Bigr]\cr
      &+&\!\!\!\!C_{-}^\uparrow(\mbox{\boldmath$\pi$})h^\uparrow_-\exp\Bigl[ i (\omega-\Omega) t\Bigr]
            +\!C_{-}^\downarrow(\mbox{\boldmath$\pi$})h^\downarrow_-\exp\Bigl[ i (\omega+\Omega) t\Bigr]\,\Biggr\}
            \exp [i(p_y y+p_z z)/\hbar] \,d^3\pi\, ,
            {\qquad}
\end{eqnarray}
where
\begin{equation}\label{Spi}
C_{+}^\uparrow(\mbox{\boldmath$\pi$})\!={\cal A}\!\left(
                  \begin{array}{c}
                    1 \\
                  0 \\
                  \!\!\!K_{2}\pi_z \!\!\! \\
                  \!\!\!K_{2}\pi_+ \!\!\!\\
                  \end{array}
                \right)\!,\ \
C_{+}^\downarrow(\mbox{\boldmath$\pi$})\!={\cal B}\!\left(
                     \begin{array}{c}
                       0 \\
                       1 \\
                     \!\!\!K_{1}\pi_- \!\!\! \\
                     \!\!\!-K_{1}\pi_z \!\!\!\\
                     \end{array}
                   \right)\!,\ \
C_{-}^\uparrow(\mbox{\boldmath$\pi$})\!={\cal C}\!\left(
                  \begin{array}{c}
                    \!\!\!-K_{1}\pi_z \!\!\! \\
                    \!\!\!-K_{1}\pi_+ \!\!\!\\
                    1 \\
                    0\\
                  \end{array}
                \right)\!,\ \
C_{-}^\downarrow(\mbox{\boldmath$\pi$})\!={\cal D}\!\left(
                  \begin{array}{c}
                    \!\!\!-K_{2}\pi_- \!\!\! \\
                    \!\!\!K_{2}\pi_z \!\!\! \\
                    0 \\
                    1\\
                  \end{array}
                \right)\!,
\end{equation}
where $\pi_{\pm}\equiv\pi_x\pm i\pi_y$. Implicitly, the
normalization condition of the wave packet~(\ref{Ansc}) is
provided if one requires that
\begin{eqnarray}\label{Norm6}
\int\left[\left|C^{\uparrow\ast}_+(\mbox{\boldmath$\pi$})\right|^2
    +\left|C^{\uparrow\ast}_-(\mbox{\boldmath$\pi$})\right|^2
    +\left|C^{\downarrow\ast}_+(\mbox{\boldmath$\pi$})\right|^2
    +\left|C^{\downarrow\ast}_-(\mbox{\boldmath$\pi$})\right|^2\right]d^3\pi=1\,.
\end{eqnarray}

In order to write the exact forms of
$C^{\uparrow\downarrow}_\pm(\mbox{\boldmath$\pi$})$, one needs to
have an explicit expression for $\Psi$, which leads to evaluation
of $<\!\dot{\mbox{\boldmath$r$}}\!>$ = c
$<\!\mbox{\boldmath$\alpha$}\!>$. For this purpose, we take the
initial state of the wave packet as a localized spin--up electron
in the $z$--direction while its center is at rest in the origin,
namely
\begin{equation}\label{Int}
\Psi(\mbox{\boldmath$r$},0)=\left(
            \begin{array}{c}
            1 \\
            0 \\
            0 \\
            0\\
            \end{array}
          \right)f\left(\frac{r}{r_{_o}}\right),
\end{equation}
where
$f(r/r_{_o})=\left[2/(\bar{\pi}r_{_o}^2)\right]^{3/4}\exp\left[-(r/r_{_o})^2\right]$
is a normalized Gaussian function with $r\equiv|r|$ and ${r_{_o}}$
as a constant that indicates approximate spatial extension of the
wave packet. The notation $\bar{\pi}$ has been introduced to
indicate the usual pi number in order not to be confused with the
generalized momentum $\mbox{\boldmath$\pi$}$. The Fourier
transformation of $f(r/r_{_o})$ is
\begin{equation}\label{Gus2}
f(\frac{\pi}{\pi_{_{o}}})=(\frac{2}{\bar{\pi}\pi_{_o}^2})^{3/4}\exp\Bigl[-(\frac{\pi}{\pi_{_o}})^2\Bigr],
\end{equation}
which, taking $\mbox{\boldmath$\pi$}$ in the spherical coordinates
with
$\mbox{\boldmath$\pi$}=\mbox{\boldmath$\pi$}(\pi,\theta,\varphi)$,
satisfies the normalization condition $\int
f^2(\pi/\pi_{_{o}})\,d^3\pi=1$, and where the constant
$\pi_{_{o}}=2\hbar/r_{_{o}}$ gives the width of wave packet in the
momentum space~\cite{huang1952}.

As we have applied the non--relativistic approximation, to
simplify the model further, we drop terms of the order
$\epsilon^2\equiv(\pi/2mc)^2$ or higher. Thus, in the weak field
approximation, $\Omega\ll\omega$, one has
$(\pi/2mc)(\hbar\Omega/2mc^2)\approx {\cal O}(\epsilon^2)$ or
higher, and then from~(\ref{Norm1}) one can write
\begin{equation}\label{Norm4}
\left(
            \begin{array}{c}
            K_1 \\
            K_2 \\
            \end{array}
          \right)\pi
\simeq\frac{\pi}{2mc}\left(
            \begin{array}{c}
            1+\frac{\hbar\Omega}{2mc^2} \\
            \\
            1-\frac{\hbar\Omega}{2mc^2} \\
            \end{array}
          \right)\simeq\frac{\pi}{2mc}\left(
            \begin{array}{c}
            1 \\
            1 \\
            \end{array}
          \right)\equiv K\pi\left(
            \begin{array}{c}
            1 \\
            1 \\
            \end{array}
          \right)\!\, .
\end{equation}
In addition, during the calculation one encounters
non--commutative terms due to the existence of the magnetic field.
For our choice of vector potential, when $\pi_i$'s are operators
(i.e. not integration variables), one gets
\begin{eqnarray}\label{commut}
\pi_+\pi_- -\pi_-\pi_+=2i(\pi_y\pi_x-\pi_x\pi_y)=\frac{2 e
B\hbar}{c}\, .
\end{eqnarray}
Therefore, up to the order of approximation we are dealing with,
the non--commutativity of $\pi_x$ and $\pi_y$ can be
ignored\rlap,\footnote{Note that, for a typical field of one
Tesla, the right hand side of relation~(\ref{commut}) is an order
of about $10^{-63}$.}\
 as assumed before, and we use $\pi_+\pi_- \simeq\pi_-\pi_+$ in the
following. Also, regarding the fact that
$h^{\uparrow\downarrow}_{\pm}(\xi)\propto H_n(\xi)\exp(
{-\xi^2/2})$, one gets $(\pi_x
h^{\uparrow\downarrow}_{\pm}-h^{\uparrow\downarrow}_{\pm}\pi_x)$
to be of the order $\hbar\sqrt{eB/c}$, where it can also be
ignored in the proposed approximation. Hence, using rough
coefficients~(\ref{Coff}) for the initial state (\ref{Int}), the
wave packet~(\ref{Ansc}) reads
\begin{eqnarray}\label{vawf}
\Psi({\mbox{\boldmath$r$}},t)\!\!\!&\simeq&\!\!\!h^{-3/2}\int\Biggl\{\left(
                 \begin{array}{c}
                   1 \\
                 0 \\
                 K\pi_z \\
                 K\pi_+ \\
                 \end{array}
               \right)\exp\Bigl[-i(\omega+\Omega)t\Bigr]+\left(
               \begin{array}{c}
                 0 \\
               0 \\
               -K\pi_z \\
               0\\
               \end{array}
             \right)\exp\Bigl[i(\omega-\Omega) t\Bigr]\cr
            & +&\!\!\!
            \left(
              \begin{array}{c}
                 0 \\
              0 \\
              0 \\
              -K\pi_+ \\
              \end{array}
            \right)\exp\Bigl[i(\omega+\Omega) t\Bigr]\Biggr\}f(\pi/\pi_{_{o}})
            \exp [i(p_y y+p_z z)/\hbar] \,d^3\pi\, .\qquad
\end{eqnarray}
Note that, each of the above rough solutions satisfy the Dirac
equation (\ref{Ham}) when the aforementioned approximations are
taken into account. Actually, by the exact
coefficients~(\ref{Coff}), they do fulfill the Dirac equation
exactly, and our approximations do~not alter the results for
altered Zbw, rather they simplify calculations.

Since the wave function must not spread rapidly in time, one can
justify the non--relativistic approximation applied
in~(\ref{Norm4}) if $r_{_{o}}$ is taken to be very large compared
to the Compton wavelength~\cite{huang1952} or equally
$\pi_{_{o}}/mc\ll1$. Therefore, all momenta which are greater than
$\pi_{_{o}}$ are diminished in the integrand~(\ref{vawf}), and
hence, the main role is played by momenta that satisfy
$\pi/mc\ll1$. This justifies the non--relativistic approximation.

Now, making use of the expectation value formula for velocity
vector,
\begin{equation}\label{Exp}
<\dot{\mbox{\boldmath$r$}}>=\int\Psi^{\ast}({\mbox{\boldmath$r$}},t)(c\mbox{\boldmath$\alpha$})
\Psi({\mbox{\boldmath$r$}},t)d^{3}r
\end{equation}
with the wave packet~(\ref{vawf}), it is straightforward to show
that for an initially localized spin--up electron
\begin{eqnarray}\label{Expv}
\!\!\!
<{\dot{\mbox{\boldmath$r$}}}>=\!\!{\mathbf{V}}\!\!\!+2c\int\Biggl\{l_{1}(\mbox{\boldmath$\pi$})
 \cos\Bigl[(\omega_{_{\rm zbw}}\!\!\!+\omega_c)t+\varphi(\mbox{\boldmath$\pi$})\Bigr]
 \!\!+l_{2}
 (\mbox{\boldmath$\pi$})\cos\,\omega_{_{\rm zbw}} t\Biggr\}\,d^{3}\pi\, ,\!\!\!
\end{eqnarray}
where
\begin{equation}\label{Lean}
{\mathbf{V}}=
c\int\left(C^{\uparrow\ast}_+\mbox{\boldmath$\alpha$}
C_{+}^\uparrow +C^{\uparrow\ast}_-\mbox{\boldmath$\alpha$}
C_{-}^\uparrow+C^{\downarrow\ast}_-
\mbox{\boldmath$\alpha$}C_{-}^\downarrow\right)d^{3}\pi
\end{equation}
is the linear velocity of the average motion of electron,
 $l_{1}(\mbox{\boldmath$\pi$})=\left|C^{\uparrow\ast}_+\mbox{\boldmath$\alpha$}
C_{-}^\downarrow\right|$,\ \,
$l_{2}(\mbox{\boldmath$\pi$})=\left|C^{\uparrow\ast}_+\mbox{\boldmath$\alpha$}
C_{-}^\uparrow\right|$
 and the phase term turns out to be
\begin{equation}\label{Phas}
\varphi(\mbox{\boldmath$\pi$})=\tan^{-1}\frac{{\rm
Im}\left(C^{\uparrow\ast}_+\mbox{\boldmath$\alpha$}C_{-}^\downarrow\right)}{{\rm
Re}\left(C^{\uparrow\ast}_+\mbox{\boldmath$\alpha$}C_{-}^\downarrow\right)
}\ .
\end{equation}
Integration of Eq.~(\ref{Expv}) in the weak field approximation
reads
\begin{eqnarray}\label{Expv1}
 <\!{\mbox{\boldmath$r$}}\!>\simeq {\mbox{\boldmath$r$}}_{_{\rm i}}\!\!+\!\!{\mathbf{V}}t+\!\!\lambdabar_c\!\!\int\!\Biggl\{l(\mbox{
 \boldmath$\pi$})
 \sin\Bigl[(\omega_{_{\rm zbw}}\!\!\!+\omega_{_{c}})t+\!\varphi(\mbox{\boldmath$\pi$})\Bigr]
 \!\!+l_{2}(\mbox{\boldmath$\pi$})\sin\,\omega_{_{\rm zbw}}t\Biggr\}\,d^{3}
\pi,
\end{eqnarray}
where $l(\mbox{\boldmath$\pi$})\simeq
l_{1}(\mbox{\boldmath$\pi$})\left(1-\omega_c/\omega_{_{\rm
zbw}}\right)$
 and we assume ${\mbox{\boldmath$r$}}_{_{\rm i}}\equiv{\mbox{\boldmath$r$}}_{\rm initial}=0$.
This relation, compared to the results with no external magnetic
field, shows that the oscillatory part -- which involves
interference between the positive and negative energy states of
electron -- is now split into two terms. As it will be discussed
bellow, within our approximation, the first term of the
oscillatory part signifies the Zbw circular motion in the
$xy$--plane with a shifted frequency, while the second term is
responsible for Zbw in the $z$--direction with an intact
frequency, as expected.

For the wave packet~(\ref{vawf}), the linear velocity term in
$<\!{\mbox{\boldmath$r$}}\!>$ vanishes. Whereas, up to the first
order in $\pi/mc$, we have
\begin{eqnarray}\label{Kfac2}
&l_{1x}=l_{1y}\simeq
-f^2(\frac{\pi}{\pi_{_{o}}})\frac{(\pi_x^2+\pi_y^2)^\frac{1}{2}}
{2mc}\, ,\qquad l_{2x}=l_{2y}= l_{1z}\simeq0\, ,\qquad
l_{2z}\simeq -f^2(\frac{\pi}{\pi_{_{o}}})\frac{\pi_z}{2mc}\, ,\cr
{}\cr
&\varphi_{x}(\mbox{\boldmath$\pi$})\simeq\tan^{-1}\left(\frac{\pi_y}{\pi_x}\right),\qquad
\varphi_{y}(\mbox{\boldmath$\pi$})\simeq\tan^{-1}\left(-\frac{\pi_x}{\pi_y}\right)\qquad
{\rm and}\qquad\varphi_{z}(\mbox{\boldmath$\pi$})=0.
\end{eqnarray}
In the spherical coordinates, one gets
\begin{eqnarray}\label{Kf}
l_{1x}=l_{1y}\simeq{-f^2(\frac{\pi}{\pi_{_{o}}})\frac{\pi\sin\theta}{2mc}}\,
,\qquad l_{2x}=l_{2y}= l_{1z}\simeq0\, ,\qquad l_{2z}\simeq
-f^2(\frac{\pi}{\pi_{_{o}}})\frac{\pi\cos\theta}{2mc}\, ,\cr
 {}\cr
\varphi_{x}(\mbox{\boldmath$\pi$})\simeq\tan^{-1}\left(\tan\varphi\right)=\varphi\,
,\qquad
\varphi_{y}(\mbox{\boldmath$\pi$})\simeq\tan^{-1}\left(-\cot\varphi\right)=
\varphi+\frac{\bar{\pi}}{2}\qquad {\rm and}\qquad
\varphi_{z}(\mbox{\boldmath$\pi$})=0.
\end{eqnarray}
Substituting these terms in relation~(\ref{Expv1}), it becomes
\begin{eqnarray}\label{Spc1}
&<x>\simeq
I^\uparrow\frac{\lambdabar_c}{2}\int_{0}^{2\bar{\pi}}\sin\Bigl[\left(\omega_{\rm
zbw}+\omega_c\right)t+\varphi\Bigr]\, d\varphi\, ,\qquad
 <y>\simeq
I^\uparrow\frac{\lambdabar_c}{2}\int_{0}^{2\bar{\pi}}\cos\Bigl[\left(\omega_{\rm
zbw}+\omega_c\right)t+\varphi\Bigr]\, d\varphi\cr {}\cr
 &
{\rm and}\qquad <z>\simeq J\lambdabar_c\sin\left(\omega_{\rm zbw}
t\right),
\end{eqnarray}
where
\begin{equation}\label{Cospc}
I^\uparrow\equiv-2\!\left(1-\!\frac{\omega_c}{\omega_{_{\rm
zbw}}}\right)\!\int_{0}^{\infty}\!\int_{0}^{\bar{\pi}}\frac{f^2(\frac{\pi}{\pi_{_{o}}})}{2mc}
\pi^3\sin^2\theta\,d\theta
\,d\pi=-(8\bar{\pi})^{-\frac{1}{2}}\frac{\lambdabar_c}{r_{_o}}\left(1-\!\frac{\omega_c}{\omega_{_{\rm
zbw}}}\right)
\end{equation}
and
\begin{eqnarray}\label{Cospc1}
J\equiv-\bar{\pi}\int_{0}^{\infty}\int_{0}^{\bar{\pi}}\frac{f^2(\frac{\pi}{\pi_{_{o}}})}{2mc}\pi^3
\sin2\theta\,d\theta\,d\pi=0.
\end{eqnarray}
As $\varphi$ is the azimuthal angle in the spherical momentum
space, the nature of the integral in the $xy$--plane over
$\varphi$ restricts us to a fixed value, e.g. $\varphi_{_{o}}$, in
order to figure out the role of a specific Fourier decomposition
in the Zbw circular motion. Of course, if one accomplishes the
integrations over the full domain, then all oscillatory motions
will disappear, and the center of the wave packet as a whole
remains at rest, as expected. Hence, by considering those Fourier
components in the $xy$--plane within $\varphi$ and
$\varphi+d\varphi$, one gets
\begin{equation}\label{Spc11}
<x>_{\varphi_{_{o}}}\simeq
\frac{\lambdabar_c}{2}\sin\Bigl[\left(\omega_{_{\rm
zbw}}+\omega_{c}\right)t+\varphi_{_{o}}\Bigr]\qquad {\rm
and}\qquad
 <y>_{\varphi_{_{o}}}\simeq
-\frac{\lambdabar_c}{2}\cos\Bigl[\left(\omega_{_{\rm
zbw}}+\omega_{c}\right)t+\varphi_{_{o}}\Bigr].
\end{equation}
The notation $I^\uparrow$ -- shown in Eq.~(\ref{Cospc}) to be
proportional to $\lambdabar_c/r_{_o}$ -- only determines the
weight of Zbw relative to the degree of localization of electron.
Eqs.~(\ref{Spc11}) express a circular motion in the $xy$--plane,
therefor one can conclude that the Zbw frequency of rotation for a
spin--up electron in an external uniform static magnetic field
changes by
\begin{equation}\label{Shiff1}
\omega^\uparrow_{_{\rm zbw}}\rightarrow\omega^\uparrow_{_{\rm
zbw}}+\omega_c\, ,
\end{equation}
while the frequency for $<\! z\!>$ remains unchanged, as expected.
The same procedure for the spin--down state yields
\begin{eqnarray}\label{Spc2}
&<x>\simeq
I^\downarrow\frac{\lambdabar_c}{2}\int_{0}^{2\bar{\pi}}\sin\Bigl[\left(\omega_{_{\rm
zbw}}-\omega_{c}\right)t-\varphi\Bigr]\, d\varphi \,,\qquad\quad
<y>\simeq
I^\downarrow\frac{\lambdabar_c}{2}\int_{0}^{2\bar{\pi}}\cos\Bigl[\left(\omega_{_{\rm
zbw}}-\omega_{c}\right)t-\varphi\Bigr]\, d\varphi\cr
 {}\cr
&{\rm and}\qquad <z>\simeq J\lambdabar_c\sin\left(\omega_{_{\rm
zbw}} t\right),
 \end{eqnarray}
with
\begin{equation}\label{Cospc2}
I^\downarrow\equiv-2\left(1+\frac{\omega_c}{\omega_{_{\rm
zbw}}}\right)\!\int_{0}^{
\infty}\!\int_{0}^{\bar{\pi}}\frac{f^2(\frac{\pi}{\pi_{_{o}}})}{2mc}
\pi^3\sin^2\theta\,d\theta d\pi
=-(8\bar{\pi})^{-\frac{1}{2}}\frac{\lambdabar_c}{r_{_o}}\left(1+\frac{\omega_c}{\omega_{_{\rm
zbw}}}\right).
\end{equation}
Thus, for a spin--down electron, the frequency changes by
\begin{equation}\label{Shiff2}
\omega^\downarrow_{_{\rm zbw}}\rightarrow\omega^\downarrow_{_{\rm
zbw}}-\omega_c\, .
\end{equation}
Obviously, the new Zbw frequencies reduce to the field free case
in the absence of external magnetic field.

Motion of electron in an external uniform static magnetic field,
employing the Heisenberg picture, has been discussed in
Ref.~\cite{Villavicencio-etal2000}. They have also shown that the
Zbw frequency, in the weak field approximation, changes to
\begin{equation}\label{UJ3}
\omega_{_{\rm zbw}}\pm\omega_c\, ,
\end{equation}
without explicitly distinguishing the spin--up from spin--down
states. However, we have proved that one can assign a distinct
modification to the Zbw frequency of each of these states given in
relations~(\ref{Shiff1}) and~(\ref{Shiff2}). Though, one may still
claim that $\omega_{_{\rm zbw}}\pm\omega_c$\ term can intuitively
be understood to stand for the spin--up/down states, respectively.
But, such an interpretation has explicitly been revealed within
our results. Besides, our outcome can be compared with the
classical results via the BZ approach~\cite{SalesiRecami2000},
that will be accomplished in the next section.

But before that, in order to gain further insight into the
situation raised by the phenomenon of Zbw, one can examine the
effect of an external uniform static magnetic field on some matrix
elements that do~not vanish. For this purpose, let us calculate
the expectation value of the magnetic moment of a spin--up/down
electron, that in the operator form it is~\cite{huang1952}
\begin{equation}\label{expmagmom}
<\!\mbox{\boldmath$\mu$}\!>=\frac{e}{2c}<\!{\mbox{\boldmath$r$}}\times\dot{\mbox{\boldmath$r$}}\!>
=\frac{e}{2}<\!{\mbox{\boldmath$r$}}\times\mbox{\boldmath$\alpha$}\!>\rightarrow
\frac{ie\hbar}{2}<\!\mbox{\boldmath$\nabla_{\pi}$}\times\mbox{\boldmath$\alpha$}\!>.
\end{equation}
Using the wave packet (\ref{vawf}), one can easily show that, with
the initial condition (\ref{Int}), the contribution from each
Fourier component of expression (\ref{expmagmom}) vanishes
individually in the $x$-- and $y$--directions, but survives in the
$z$--direction, as maybe expected. That is, from classical point
of view, the $z$--component of (\ref{expmagmom}), which is
perpendicular to the circle of Zbw motion, is the only
non--vanishing component of the magnetic moment. Straightforward
calculations, using the employed approximations, reveal that one
gets
\begin{equation}\label{expmagmomup}
<\!\mu^\uparrow_x\!>=0,\qquad<\!\mu^\uparrow_y\!>=0\qquad{\rm
and}\qquad <\!\mu^\uparrow_z\!>=-\frac{\mid\!
e\!\mid\lambdabar_c}{2}\biggl[1-\cos(\omega_{_{\rm
zbw}}+\omega_{c})t\biggr]
\end{equation}
for the spin--up states, and
\begin{equation}\label{expmagmomdown}
<\!\mu^\downarrow_x\!>=0,\qquad<\!\mu^\downarrow_y\!>=0\qquad{\rm
and}\qquad <\!\mu^\downarrow_z\!>=\frac{\mid\!
e\!\mid\lambdabar_c}{2}\biggl[1-\cos(\omega_{_{\rm
zbw}}-\omega_{c})t\biggr]
\end{equation}
for the spin--down states. These results indicate that each
Fourier wave contributes a circular motion about the direction of
spin. Hence, one may again interpret that the intrinsic magnetic
moment is a result of the (shifted) Zbw. Incidentally, the time
dependant part of the magnetic moment, in
relations~(\ref{expmagmomup}) and~(\ref{expmagmomdown}), is a
consequence of the fact that we have assumed that the spin of
electron had initially been observed. This requirement leads to
bring in the negative energy parts which do~not vanish by letting
the wave packet spread in space~\cite{huang1952}.

Furthermore, it would be instructive to employ the classical
interpretation of
spin~\cite{huang1952,hestenes93,barut-etal81b,hestenes90,hestenes2009}
via the intrinsic/microscopic angular momentum vector caused from
the current produced by the local Zbw frequency when this
frequency is shifted in the presence of an external magnetic
field. For this purpose, by relations~(\ref{Spc1})
and~(\ref{Spc2}) when considering~(\ref{Cospc})
and~(\ref{Cospc2}), one can claim that the radius of the Zbw
circle is $r_{_{\rm zbw}}=\lambdabar_c(1\pm\varepsilon)/2$, where
$\varepsilon\equiv -\omega_c/\omega_{_{\rm zbw}}$. And thus, the
speed of electron around the center of mass will be
\begin{equation}
v_{_{\rm
zbw}}=\left[\frac{\lambdabar_c}{2}(1\pm\varepsilon)\right]\left[\omega_{_{\rm
zbw}}(1\mp\varepsilon)\right]=c(1-\varepsilon^2)\simeq c\, .
\end{equation}
This result is consistent with the calculation of Zbw velocity via
the Heisenberg picture, where the position operator
${\mbox{\boldmath$r$}}$ satisfies $\dot{\mbox{\boldmath$r$}}=i[H,
{\mbox{\boldmath$r$}}]/\hbar=c{\mbox{\boldmath$\alpha$}}$. Hence,
the intrinsic/microscopic angular momentum is
\begin{equation}
s_{_{\rm zbw}}=r_{_{\rm zbw}}(mv_{_{\rm zbw}})\simeq
\frac{\hbar}{2}(1\pm\varepsilon)\simeq\frac{\hbar}{2(1\mp\varepsilon)}\,
.
\end{equation}
This indicates that the gyromagnetic $g$ factor of electron in the
presence of an external magnetic field changes to
$2(1\mp\varepsilon)$, however the magnetic moment, relations
(\ref{expmagmomup}) and (\ref{expmagmomdown}), shows that the time
average of the gyromagnetic $g$ factor in a period does~not
change.

On the other hand, if the spin of electron must remain unchanged,
i.e. with the fixed value of $\hbar/2$, then by assuming $v_{_{\rm
zbw}}\propto c$, (e.g. $v_{_{\rm zbw}}=\zeta c$), the radius of
the Zbw circle will be
\begin{equation}
r_{_{\rm zbw}}=\frac{v_{_{\rm zbw}}}{\omega_{_{\rm
zbw}}(1\mp\varepsilon)}=\frac{\lambdabar_c}{2}\left(\frac{\zeta}{1\mp\varepsilon}\right).
\end{equation}
Hence, the intrinsic/microscopic angular momentum is
\begin{equation}
s_{_{\rm
zbw}}=\frac{\lambdabar_c}{2}\left(\frac{\zeta}{1\mp\varepsilon}\right)(mv_{_{\rm
zbw}})
=\frac{\hbar}{2}\left(\frac{\zeta^2}{1\mp\varepsilon}\right),
\end{equation}
which if it is supposed to remain $\hbar/2$, one will get
$\zeta\simeq 1\mp\varepsilon/2$. That is, in this view, $v_{_{\rm
zbw}}\simeq c(1\mp\varepsilon/2)$ and $r_{_{\rm
zbw}}\simeq\lambdabar_c(1\pm\varepsilon/2)/2$.

\section{Frequency Shift via Classical Spinning Electron Approach}
\indent

As mentioned in the introduction, BZ
devised~\cite{SalesiRecami2000,BarutZanghi84,RecamiSalesi1995} a
new classical analogous model for the relativistic spinning
electron in the ordinary space--time by introducing a classical
Lagrangian as
\begin{equation}\label{Lag}
L=\frac{1}{2}i\alpha(\dot{\bar{z}}z-\bar{z}\dot{z})+
p_{\mu}(\dot{x}^\mu-\bar{z}\gamma^\mu z)+eA_{\mu}\bar{z}\gamma^\mu
z\, ,
\end{equation}
where $\mu=0, 1, 2, 3$, $A^\mu$ is the electromagnetic potential,
$\alpha$ is a constant with the dimension of action, $z$ and
$\bar{z}\equiv z^\dag\gamma^0$ are ordinary ${\mathcal{C}}^4$
bi--spinors which used as a second pair of conjugate classical
spinorial variables, the speed of light, $c$, is set equal to one
and dot indicates derivation with respect to the proper time,
$\tau$, where $d/d\tau=\dot{x}^\mu\partial_\mu$\,.

The four Euler--Lagrange equations can be obtained with respect to
variables $x^\mu$, $p^\mu$, $z$ and $\bar{z}$. By assuming
$-\alpha=1=\hbar$ and adopting independent dynamical variables
$x^\mu$, $\pi^\mu$, $v^\mu$ and $S^{\mu\nu}$, where $v^\mu$ is the
$4$--velocity and $S^{\mu\nu}\equiv i\bar{z}[\gamma^\mu,
\gamma^\nu]z/4$ is the {\it spin tensor} met in the Dirac theory,
the Euler--Lagrange equations are obtained within the classical
approach to be~\cite{SalesiRecami2000}
\begin{equation}\label{Magn0}
\dot{\pi}^\mu=eF^{\mu\nu}v_\nu\, ,\qquad\dot{x}^\mu=v^\mu\,
,\qquad\dot{v}^\mu=4S^{\mu\nu}\pi_\nu\qquad{\rm
and}\qquad\dot{S}^{\mu\nu}=v^\nu\pi^\mu-v^\mu\pi^\nu .
\end{equation}

For a free electron (i.e. $A^{\mu}=0$), the equation of motion for
$4$--velocity is derived to be~\cite{RecamiSalesi1995}
\begin{equation}\label{Moeq}
\mbox{$\upsilon$}^\mu=\frac{{p}^\mu}{m}
-\frac{\ddot{\mbox{$\upsilon$}}^\mu}{4m^2}\ .
\end{equation}
A general solution for this case has been shown to
be~\cite{barut-etal81a,BarutZanghi84,Barut-pavsic}
\begin{equation}\label{Velo}
\upsilon^{\mu}=\frac{p^\mu}{m}+\left[\upsilon^\mu(0)-\frac{p^\mu}{m}\right]\cos(\omega_{_{\rm
zbw}}\tau)+\frac{\dot{\upsilon}^\mu(0)}{2m}\sin(\omega_{_{\rm
zbw}}\tau)\, .
\end{equation}
The first term of this solution belongs to the rectilinear part of
the velocity, the second and third ones indicate the Zbw motion.

For a spinning electron in an external electromagnetic field,
third equation of Eqs.~(\ref{Magn0}) indicates that the general
solution represents a {\it helical} motion in the ordinary
$3$--space. In fact, the equation of motion from
Eqs.~(\ref{Magn0}) is derived to be~\cite{SalesiRecami2000}
\begin{equation}\label{Magn}
\ddot{\upsilon}^\mu-4m\pi^\mu+4m^2\upsilon^\mu+{4e\upsilon^\mu}F_{\alpha\beta}S^
{\alpha\beta}-4eS^{\mu\nu}F_{\nu\beta}\upsilon^\beta=0\, ,
\end{equation}
where $F^{\mu\nu}$\ is the electromagnetic tensor. Writing the
$x$-- and $y$--components of Eq.~(\ref{Magn}) for an electron with
the $z$--component of spin in an external uniform static magnetic
field along the $z$--direction (in which the only nonzero
components of the electromagnetic tensor are $F^{21}=-F^{12}=B$),
then deriving with respect to $\tau$ while using the first
equation of~(\ref{Magn0}) results in~\cite{SalesiRecami2000}
\begin{eqnarray}\label{Shif22unew2eqs}
\thrdotovervx+4\left(m^2-3\,es_zB\right)\dot{v}_x-4\,meBv_y=0\,,\cr
\thrdotovervy+4\left(m^2-3\,es_zB\right)\dot{v}_y+4\,meBv_x=0\,.
\end{eqnarray}
These equations describe a uniform circular motion in the
$xy$--plane whose angular velocities must satisfy the algebraic
equation~\cite{SalesiRecami2000}
\begin{equation}\label{Freq}
\omega^3-4\left(m^2-3\,es_zB\right)\omega+4\,meB=0\, ,
\end{equation}
or equivalently, with our notations,
\begin{equation}\label{NewFreq}
\omega^3-\omega_{_{\rm zbw}}^2\omega\pm
3\,\varepsilon\omega_{_{\rm zbw}}^2\omega+\varepsilon\omega_{_{\rm
zbw}}^3=0\, ,
\end{equation}
where the upper (lower) sign represents the spin--up (down) state.
Solutions of~(\ref{NewFreq}), by the weak field approximation and
up to the first--order terms in $\varepsilon $, have been asserted
to be~\cite{SalesiRecami2000}\footnote{Again, we have rewritten
their solutions with our notations.}
\begin{equation}\label{Omchr}
\omega_{_{1}}\simeq\omega_c\left(1\pm 3\,\varepsilon\right)\,
,\qquad \omega_{_{2}}\simeq\omega_{_{\rm
zbw}}\left(1\mp\frac{3}{2}\,\varepsilon\right)\qquad\textrm{and}\qquad
\omega_{_{3}}\simeq-\omega_{_{\rm
zbw}}\left(1\mp\frac{3}{2}\,\varepsilon\right).
\end{equation}

Frequency $\omega_{_{1}}$ has been
regarded~\cite{SalesiRecami2000} as the usual rotational frequency
of electron in an external magnetic field, which due to the Zbw,
differs from that of spinless particles. Namely, it shows the
effect of spin on the cyclotron frequency. It has also been
claimed~\cite{SalesiRecami2000} that the small predicted deviation
to the cyclotron frequency can be detected by their suggested
device. This argument nearly explains the purpose of
Ref.~\cite{SalesiRecami2000}. Meanwhile, the other two solutions,
$\omega_{_{2}}$ and $\omega_{_{3}}$, have been described as
modified forms of $\omega_{_{\rm zbw}}$ which, due to the
smallness of $\varepsilon$, have essentially been
considered~\cite{SalesiRecami2000} identical to $\omega_{_{\rm
zbw}}$ without further investigation.

In this work, we purpose to probe these terms more closely.
Actually, we are interested in investigating effects of external
magnetic fields on the Zbw frequency itself. Of course, there is
no doubt that due to the smallness of variations involved in this
issue, one has little hope of detection by the current
experiments, unless innovative device will
emerge\rlap.\footnote{Interested readers in related experimental
techniques may consult Ref.~\cite{Van-DyckSchwinbergDehmelt1986}
(and references therein), where they have measured the electron's
$g$ factor. Also, see
Refs.~\cite{rusin-etal2006,hestenes2009},~\cite{ZCetalFetalZJetalZZSetalK6}--\cite{LamataLeonSchatzSolano}
for more recent theoretical studies for experimentally observing
and/or experimental devices on the phenomenon of Zbw. For example,
an optical lattice scheme, which may permit experimental
observation of the Zbw with ultra--cold neutral atoms, has been
proposed in Ref.~\cite{VaishnavClark2008}, in where they also
claim that the phenomenon of Zbw will occur at experimentally
accessible frequencies and amplitudes.}\
 Examples fall short of the level of this effect, even detection of the
$\omega_{_{\rm zbw}}$ itself for free electron are outside of the
current experimental and technological
reach~\cite{rusin-etal2006,VaishnavClark2008,LamataLeonSchatzSolano}.
Unless, e.g., they are indirectly in the response of electron to
external fields and/or a kind of condition (e.g. a resonance
condition) employed to minimize the Zbw frequency. For example,
the Darwin term in the Hamiltonian of an atomic electron comes
from the Zbw motion which makes electron sensitive to the
potential of the nuclei in its average
position~\cite{Dirac58}--\cite{Merzbacher98}. Furthermore, to
grasp an intuitive sense, one may recall that the Zbw frequency of
oscillation is too high and hence the period of the time involved
is too low, whereas the time measurement sensitivity limits are
generally known to be about $10^{-16} {\rm
s}$~\cite{timeprecision}.

Nevertheless, following our purpose while probing the solutions
$\omega_{_{2}}$ and $\omega_{_{3}}$, we have noticed that these
two solutions of~(\ref{Omchr}), are very rough approximations.
Actually, the more accurate solutions $\omega_{_{2}}$ and
$\omega_{_{3}}$, after solving equation (\ref{NewFreq}), are
\begin{eqnarray}\label{Shif2}
\omega_{_{2}}\simeq\omega_{_{\rm zbw}}({{1-2\varepsilon}\atop
{1+\varepsilon}})\ \quad\qquad \textrm{and}\ \quad\qquad
\omega_{_{3}}\simeq-\omega_{_{\rm zbw}}({{1-\varepsilon}\atop
{1+2\varepsilon}}),
\end{eqnarray}
where again the upper (lower) value represents solution for the
spin--up (down) state. These solutions contradict with the
solutions derived from the quantum approaches, i.e. the approach
employed in this work via the expectation value and the one used
in Ref.~\cite{Villavicencio-etal2000} via the Heisenberg picture.

The more accurate solutions (\ref{Shif2}) lead to some
discrepancies. In fact, regarding solution~(\ref{Velo}) and via
the BZ spinning particle model, Salesi and Recami have associated
$(-\omega_{_{\rm zbw}})$ with antiparticles and $\omega_{_{\rm
zbw}}$ with particles~\cite{SalesiRecami1997}. This way, they have
concluded that the only change which happens to the Zbw part of
the motion via $\omega_{_{\rm zbw}}\rightarrow-\omega_{_{\rm
zbw}}$ would be conversion of counterclockwise motion to the
clockwise one. It means that solution~(\ref{Velo}) should treat
electrons and positrons, other than assigning a different
direction of rotation to their Zbw motions, in the same spin
states. In this merit, it has also been
inferred~\cite{SalesiRecami2000} that $\omega_{_{2}}$ and
$\omega_{_{3}}$ are generalized features of $\omega_{_{\rm zbw}}$
and $(-\omega_{_{\rm zbw}})$, respectively.

Now, let us investigate this issue for the
solutions~(\ref{Shif2}), in which they can be rewritten for
electron and positron with the spin--up as\footnote{Note that,
according to our convention $\varepsilon$ has a negative value.}
\begin{eqnarray}\label{Shif22unew}
 \omega^\uparrow_{_{\rm
zbw}}&\rightarrow&\omega^\uparrow_{_{2}}\simeq\omega^\uparrow_{_{\rm
zbw}}-2\,\varepsilon\omega^\uparrow_{_{\rm zbw}}\cr {}\cr
 (-\omega^\uparrow_{_{\rm
zbw}})&\rightarrow&\omega^\uparrow_{_{3}}\simeq
(-\omega^\uparrow_{_{\rm
zbw}})-\varepsilon(-\omega^\uparrow_{_{\rm zbw}})\, ,
\end{eqnarray}
while with the spin--down as
\begin{eqnarray}\label{Shif22dnew}
\omega^\downarrow_{_{\rm
zbw}}&\rightarrow&\omega^\downarrow_{_{2}}\simeq\omega^\downarrow_{_{\rm
zbw}}+\varepsilon\omega^\downarrow_{_{\rm zbw}}\cr {}\cr
 (-\omega^\downarrow_{_{\rm
zbw}})&\rightarrow&\omega^\downarrow_{_{3}}\simeq
(-\omega^\downarrow_{_{\rm
zbw}})+2\,\varepsilon(-\omega^\downarrow_{_{\rm zbw}})\, .
\end{eqnarray}
These relations show that the existence of an external magnetic
field breaks the usual symmetries which hold for free particles
and antiparticles. That is, the spin--up electron experiences a
new Zbw frequency while the one for the spin--up positron is
different, and vice versa for the spin--down states. In addition,
the Zbw frequency of an electron (positron) exposed to an external
magnetic field behaves oddly, and the absolute value of frequency
shift depends on the orientation of spin. Indeed, the Zbw
frequency of the spin--up particle changes twice that of the
spin--down, and vice versa for antiparticles. Therefore, the usual
symmetry of free particles and antiparticles states, which is
valid in the electromagnetic interactions, is broken. More
obvious, it means that the well--known CP symmetry, which is
respected in the quantum electrodynamics\rlap,\footnote{See
standard textbooks on the quantum field theory, e.g.
Ref.~\cite{ItzyksonZuber}. }\
 is violated, however, the quantum mechanical symmetries
might not necessarily hold in the classical level.

Incidentally, if one repeats the calculations of Sec.~$2$ with the
same initial state but for positron instead of electron, namely
\begin{equation}\label{Intposit}
\Psi({\mathbf{r}},0)=\left(
            \begin{array}{c}
            0 \\
            0 \\
            1 \\
            0\\
            \end{array}
          \right)f\left(\frac{r}{r_{_o}}\right)
\end{equation}
instead of relation~(\ref{Int}), one will get
\begin{equation}\label{Shiff1posit}
\omega^\uparrow_{_{\rm zbw}}\rightarrow\omega^\uparrow_{_{\rm
zbw}}-\omega_c
\end{equation}
and
\begin{equation}\label{Shiff2posit}
\omega^\downarrow_{_{\rm zbw}}\rightarrow\omega^\downarrow_{_{\rm
zbw}}+\omega_c
\end{equation}
instead of shifts~(\ref{Shiff1}) and~(\ref{Shiff2}). Now, if one
can justify -- e.g. based on the initial states~(\ref{Int})
and~(\ref{Intposit}) -- the application of the classical
interpretation of Ref.~\cite{SalesiRecami1997} for the expectation
value approach, then one will easily be able to show that this
approach holds the above discussed
symmetries\rlap.\footnote{However, one cannot make such a
classical energy separation within the relativistic quantum
approaches, see, e.g., the wave packet~(\ref{vawf}).}\
 In another words, our results, relations (\ref{Shiff1}), (\ref{Shiff2}),
(\ref{Shiff1posit}) and (\ref{Shiff2posit}), indicate that the
amount of frequency shifts are opposite in the spin--up/down
states and in electron/positron cases, which are consistent with
the CP symmetry.

\section{Discussion and Conclusion}
\indent

Much effort in the literature has been devoted to putting forward
a consistent theory to describe the spin of electron. Among them,
Huang~\cite{huang1952} used the Dirac equation to tackle this
issue and BZ~\cite{BarutZanghi84} introduced a new classical
model, which is believed to be the most satisfactory picture of a
classical spinning electron constituting a natural classical limit
of the Dirac equation~\cite{SalesiRecami2000,Barut-pavsic}.

In this work, we have employed the expectation value quantum
approach~\cite{huang1952} to investigate the effects of an
external magnetic field on the Zbw frequency of electron. We have
found that, in the non--relativistic weak field approximation, the
Zbw frequency in the presence of an external uniform static
magnetic field changes according to relations~(\ref{Shiff1})
and~(\ref{Shiff2}) for electrons, and
relations~(\ref{Shiff1posit}) and~(\ref{Shiff2posit}) for
positrons. Actually, the individual Fourier components of the
position expectation value of an electron with the certain initial
conditions do~not vanish, and indeed lead to an interpretation in
terms of a shifted Zbw frequency. Furthermore, not only each of
the Fourier wave contributes a circular motion in the plane
perpendicular to the direction of electron spin with radius
$\lambdabar_c/2$, rather the shifted Zbw frequency depends equally
on the orientation of the direction as well, i.e. the spin--up or
down. Our results are exactly the same as those derived in
Ref.~\cite{Villavicencio-etal2000} via the Heisenberg picture.
However, contrary to Ref.~\cite{Villavicencio-etal2000}, we have
explicitly shown the distinct modifications for the spin--up and
spin--down states, as one may intuitively expect. Besides, in
order to compare the classic and quantum approaches to the Zbw in
an external magnetic field, this characteristic of our results are
more appropriate.

The BZ classical approach has been
generalized~\cite{SalesiRecami2000} to include the case of an
electron in the presence of an external uniform static magnetic
field. Hence, it is shown that in the weak field approximation,
the motion of electron is described by the three characteristic
frequencies, where one of them reflects the effect of spin on the
cyclotron frequency and the other two are the modified forms of
$\omega_{_{\rm zbw}}$. We have noticed that the latter two derived
frequencies are very rough approximations. Thus, we have indicated
the more accurate solutions, solutions~(\ref{Shif2}). However,
contrary to the quantum approaches -- the approach employed in
this work via the expectation value and the one used in
Ref.~\cite{Villavicencio-etal2000} via the Heisenberg picture --
the solutions~(\ref{Shif2}) are controversial and break the usual
symmetry of free particles and antiparticles states. We have made
this point more obvious by writing the solutions~(\ref{Shif2}) via
the BZ~\cite{SalesiRecami1997} spinning particle and antiparticle
interpretation. Indeed, these solutions,
relations~(\ref{Shif22unew}) and~(\ref{Shif22dnew}), indicate that
the frequency shifts depend nontrivially on the orientation of
spin and also on whether one is dealing with particles or
antiparticles. Hence, CP violation occurs in the classical
electrodynamics domain. Therefore, regarding the Zbw frequency of
electron, one may conclude that the BZ classical approach in the
presence of an external magnetic field is unlikely to correctly
describe the spin of electron, while the quantum approach does.
That is, the quantum approach respects the CP symmetry, as
expected.

\setcounter{equation}{0}
\renewcommand{\theequation}{A.\arabic{equation}}
\section*{Appendix}
\indent

Using the initial state~(\ref{Int}) and
approximations~(\ref{Norm4}),~(\ref{commut}) and $(\pi_x
h^{\uparrow\downarrow}_{\pm}\simeq
h^{\uparrow\downarrow}_{\pm}\pi_x)$, the exact coefficients in
relation~(\ref{Spi}) are found roughly to be
\begin{eqnarray}\label{Coff}
 {\cal
 A}\!\!&=&\!\!\frac{h^\uparrow_-h^\downarrow_+h^\downarrow_-+h^\uparrow_-
              h^\downarrow_+h^\downarrow_-K^2_1\pi^2_z+h^\downarrow_-K_1K_2\pi_-h^\downarrow_+\pi_+h^\uparrow_-}
              {\Gamma}f(\pi/\pi_{_{o}})\simeq\frac{f(\pi/\pi_{_{o}})}{h^\uparrow_+}\ ,\cr
 {\cal
 B}\!\!&=&\!\!\frac{-h^\downarrow_-K^2_2\pi_+h^\uparrow_-\pi_zh^\uparrow_+
              +h^\uparrow_-h^\downarrow_-K_1K_2\pi_z\pi_+h^\uparrow_+}{\Gamma}f(\pi/\pi_{_{o}})\simeq 0
              \ ,\cr
 {\cal
 C}\!\!&=&\!\!-\frac{K^2_1K_2\pi_-h^\downarrow_+\pi_zh^\downarrow_-\pi_+h^\uparrow_+
              +h^\uparrow_+h^\downarrow_+h^\downarrow_-K^2_1K_2\pi^3_z
              +h^\uparrow_+h^\downarrow_+h^\downarrow_-K_2\pi_z}{\Gamma}
              f(\pi/\pi_{_{o}})\cr
   \!\!&\simeq &\!\!\frac{-K\pi_z}{h^\uparrow_-}f(\pi/\pi_{_{o}})\ ,\cr
 {\cal
 D}\!\!&=&\!\!-\frac{h^\uparrow_+h^\downarrow_+K_1K^2_2\pi^2_z\pi_+h^\uparrow_-
              +K_1K^2_2\pi_+h^\uparrow_+\pi_-h^\downarrow_+\pi_+h^\uparrow_-
              +h^\uparrow_-h^\downarrow_+K_2\pi_+h^\uparrow_+}{\Gamma}f(\pi/\pi_{_{o}})\cr
   \!\!&\simeq &\!\!\frac{-K\pi_+}{h^\downarrow_-}f(\pi/\pi_{_{o}})\ ,
\end{eqnarray}
where
\begin{eqnarray}\label{gamma}
\Gamma\!\!&\equiv&\!\!h^\uparrow_+h^\uparrow_-h^\downarrow_+h^\downarrow_-
                      +h^\uparrow_+h^\downarrow_+K_1^2K_2^2\pi_z^2\pi_-h^\downarrow_-\pi_+h^\uparrow_-
                      +h^\uparrow_-h^\downarrow_-K_1^2K_2^2\pi_z^2\pi_-h^\downarrow_+\pi_+h^\uparrow_+\cr
      &&\!\!\!\!+h^\uparrow_+h^\uparrow_-h^\downarrow_+h^\downarrow_-K_1^2\pi_z^2
        +h^\uparrow_+h^\uparrow_-h^\downarrow_+h^\downarrow_-K_1^2K_2^2\pi_z^4
        +h^\uparrow_+h^\uparrow_-h^\downarrow_+h^\downarrow_-K_2^2\pi_z^2\cr
      &&\!\!\!\!+h^\uparrow_+h^\downarrow_-K_1K_2\pi_-h^\downarrow_+\pi_+h^\uparrow_-
        +K_1^2K_2^2\pi_-h^\downarrow_+\pi_-h^\downarrow_-\pi_+h^\uparrow_-\pi_+h^\uparrow_+
        +h^\uparrow_-h^\downarrow_+K_1K_2\pi_+h^\uparrow_+\pi_-h^\downarrow_-\, .
\end{eqnarray}

\section*{Acknowledgement}
\indent

We thank the Research Office of the Shahid Beheshti University for
financial support. M.Z.--A. would like to appreciate his
supervisor M.F. for very useful comments which have made the
completion of this work possible.

%

\begin{thebibliography}{9}
\bibitem{dirac2-1928}P.A.M. Dirac, {\it Proc. Roy. Soc.}\ {\bf A117} (1928), 610-624; {\it ibid}\ {\bf A118} (1928), 351-361.
\bibitem{Dirac58}P.A.M. Dirac, {\it The Principles of Quantum Mechanics},
                                                        (Oxford University Press, Oxford, 4th ed. 1958).
\bibitem{Rose61}M.E. Rose, {\it Relativistic Electron Theory},
                                                        (John Wiley \& Sons, New York, 1961).
\bibitem{BjorkenDrell64}J.D. Bjorken \& S.D. Drell, {\it Relativistic Quantum Mechanics},
                                                        (McGraw--Hill, New York, 1964).
\bibitem{Sakurai67}J.J. Sakurai, {\it Advanced Quantum Mechanics},
                                                        (Pearson Education, Inc., Delhi India, 1967).
\bibitem{ItzyksonZuber}C. Itzykson \& J.-B. Zuber, {\it Quantum Field Theory}, (McGraw--Hill, Singapore, 1980).
\bibitem{Greiner98}W. Greiner, {\it Relativistic Quantum Mechanics}, (Springer--Verlag, Berlin, 1990).
\bibitem{Merzbacher98}E. Merzbacher, {\it Quantum Mechanics},
                                                        (John Wiley \& Sons, New York, 3rd ed. 1998).
\bibitem{huang1952}K. Huang, {\it Am. J. Phys.}\ {\bf 20} (1952), 479-484.
\bibitem{schrodinger30-31}E. Schr\"odinger, {\it Sitz. Preuss. Akad. Wiss. Phys. Math. Kl.}\ {\bf 24} (1930), 418-428;
                          {\it ibid}\ {\bf 3} (1931), 1.
\bibitem{barut-etal81a}A.O. Barut \& A.J. Bracken, {\it Phys. Rev.}\ {\bf D23} (1981), 2454-2463.
\bibitem{newton-etal49}T.D. Newton \& P. Wigner, {\it Rev. Mod. Phys.}\ {\bf 21} (1949), 400-406.
\bibitem{rusin-etal2006}T.M. Rusin \& W. Zawadzki, {\it J. Phys.: Cond. Matter}\ {\bf 19} (2007), 136219 [18 pages];
                        {\it Phys. Rev.}\ {\bf B76} (2007), 195439 [7 pages].
\bibitem{bender84}D. Bender \& {\it et al.}, {\it Phys. Rev.}\ {\bf D30} (1984), 515-527.
\bibitem{hestenes93}D. Hestenes, {\it Found. Phys.}\ {\bf 23} (1993), 365-387.
\bibitem{gordon27frenkel341}W. Gordon, {\it Z. Physik}\ {\bf 50} (1927), 630-632.
\bibitem{gordon27frenkel342}J. Frenkel, {\it Wave Mechanics, Advanced General Theory}, (Oxford University Press, London, 1934).
\bibitem{barut-etal81b}A.O. Barut \& A.J. Bracken, {\it Phys. Rev.}\ {\bf D24} (1981), 3333-3334.
\bibitem{hestenes90}D. Hestenes, {\it Found. Phys.}\ {\bf 20} (1990), 1213-1232.
\bibitem{hestenes2009}D. Hestenes, {\it Found. Phys.}\ {\bf 40} (2010), 1-54.
\bibitem{MHWB1}M. Mathissen, {\it Acta Phys. Polon}\ {\bf 6} (1937), 163-200; {\it ibid} (1937), 218.
\bibitem{MHWB2}H. H\"onl, {\it Ann. Physik}\ {\bf 33} (1938), 565-585.
\bibitem{MHWB3}J.W. Weyssenhof, {\it Nature}\ {\bf 141} (1938), 328-329; {\it Acta Phys. Polon}\ {\bf 9} (1947), 47-53.
\bibitem{MHWB4}J. Brandm\"uller, {\it Naturwiss.}\ {\bf 38} (1951), 139-139.
\bibitem{SalesiRecami2000}G. Salesi \& E. Recami, {\it Phys. Lett.}\ {\bf A267} (2000), 219-224.
\bibitem{PFetalT1}M.H.L. Pryce, {\it Proc. Roy. Soc. London}\ {\bf A195} (1948), 62-81.
\bibitem{PFetalT2}L.L. Foldy \& S.A. Wouthuysen, {\it Phys. Rev.}\ {\bf 78} (1950), 29-36.
\bibitem{PFetalT3}S. Tani, {\it Progr. Theor. Phys.}\ {\bf 6} (1951), 267-285.
\bibitem{FetalBetalL1}H. Feshbach \& F. Villars, {\it Rev. Mod. Phys.}\ {\bf 30} (1958), 24-25.
\bibitem{FetalBetalL2}A.O. Barut \& S. Malin, {\it Rev. Mod. Phys.}\ {\bf 40} (1968), 632-651.
\bibitem{FetalBetalL3}J.A. Lock, {\it Am. J. Phys.}\ {\bf 47} (1979), 797-802.
\bibitem{Krekora-etal2004}P. Krekora, Q. Su \& R. Grobe,  {\it Phys. Rev. Lett.}\ {\bf 93} (2004), 043004 [4 pages].
\bibitem{BarutZanghi84}A.O. Barut \& N. Zanghi, {\it Phys. Rev. Lett.}\ {\bf 52} (1984), 2009-2012.
\bibitem{RecamiSalesi98}E. Recami \& G. Salesi, {\it Phys. Rev.}\ {\bf A57} (1998), 98-105.
\bibitem{Salesi2002}G. Salesi, {\it Int. J. Mod. Phys.}\ {\bf A17}
                    (2002), 347-374.
\bibitem{Barut-pavsic}A.O. Barut \& M. Pav\v{s}i\v{c}, {\it Class. Quant. Grav.}\ {\bf 4} (1987), L131-L136.
\bibitem{RecamiSalesi1995}E. Recami \& G. Salesi, ``Field theory
                          of the electron: spin and Zitterbewegung'', in {\it Particles, Gravity and
                          Space--Time}, Ed. P.I. Pronin \& G.A. Sardanashvily (World Scientific, Singapore, 1996),
                          pp. 345-368, [hep-th/9508168].
\bibitem{Pavsic-elal1998}M. Pav\v{s}i\v{c}, E. Recami \& W.A.
                         Rodrigues Jr., {\it Hadronic J.}\ {\bf 18} (1995), 97-118.
\bibitem{ZCetalFetalZJetalZZSetalK1}W. Zawadzki, in {\it Optical Properties of Solides}, Ed.
                                    E.D. Haidemenakis (Gordon \& Breach, New York, 1970), p. 179.
\bibitem{ZCetalFetalZJetalZZSetalK3}F. Cannata, L. Ferrari \& G. Russo, {\it Solid State Commun.}\ {\bf 74} (1990), 309-312.
\bibitem{ZCetalFetalZJetalZZSetalK4}L. Ferrari \& G. Russo, {\it Phys. Rev.}\ {\bf B42} (1990), 7454-7461.
\bibitem{ZCetalFetalZJetalZZSetalK5}W. Zawadzki, in {\it High Magnetic Fields in the Physics of Semiconductors II},
                                    Ed. G. Landwehr \& W. Ossau (World Scientific, Singapore, 1997), p. 755.
\bibitem{Salesi2005}G. Salesi, {\it Int. J. Mod. Phys.}\ {\bf A20}
                    (2005), 2027-2036.
\bibitem{ZCetalFetalZJetalZZSetalK6}Z.F. Jiang, R.D. Li, S.C. Zhang \& W.M. Liu, {\it Phys. Rev.}\ {\bf B72} (2005), 045201 [5 pages].
\bibitem{ZCetalFetalZJetalZZSetalK7}W. Zawadzki, {\it Phys. Rev.}\ {\bf B72} (2005), 085217 [4 pages].
\bibitem{ZCetalFetalZJetalZZSetalK8}W. Zawadzki, ``One--dimensional semirelativity for electrons in carbon nanotubes'',
                                    {\it cond-mat/0510184} [4 pages].
\bibitem{ZCetalFetalZJetalZZSetalK9}J. Schliemann, D. Loss \& R.M. Westervelt, {\it Phys. Rev. Lett.}\
                                     {\bf 94} (2005), 206801 [4 pages]; {\it Phys. Rev.}\ {\bf B73} (2006), 085323 [9 pages].
\bibitem{ZCetalFetalZJetalZZSetalK10}M.I. Katsnelson, {\it Eur. Phys. J.}\ {\bf B51} (2006), 157-160.
\bibitem{CsertiDavid2006}J. Cserti \& G. D\'{a}vid, {\it Phys. Rev.}\ {\bf B74} (2006),
                         172305 [4 pages].
\bibitem{VaishnavClark2008}J.Y. Vaishnav \& C.W. Clark, {\it Phys. Rev. Lett.}\
                           {\bf 100} (2008), 153002 [4 pages].
\bibitem{BrusheimXu2008}P. Brusheim \& H.Q. Xu, ``Catching the Zitterbewegung'', {\it con-mat/0810.2186} [4 pages].
\bibitem{LamataLeonSchatzSolano}L. Lamata, J. Le\'{o}n, T. Sch\"{a}tz \& E. Solano, {\it Phys. Rev. Lett.}\
                                {\bf 98} (2007), 253005 [4 pages].
\bibitem{Villavicencio-etal2000}M. Villavicencio \& J.A.E. Roa--Neri, {\it Eur. J. Phys.}\ {\bf 21} (2000), 119-123.
\bibitem{holten92}J.W. van Holten, {\it Physica A.}\ {\bf 182} (1992), 279-292.
\bibitem{holten93}J.W. van Holten, ``Relativistic dynamics of spin in strong external fields'', {\it hep-th/9303124} [11 pages].
\bibitem{thirring58}H. Thirring, {\it Principles of Quantum Electrodynamics}, (Academic Press, New York, 1958).
\bibitem{Van-DyckSchwinbergDehmelt1986}R.S. Van Dyck Jr., P.B. Schwinberg \& H.G. Dehmelt, {\it Phys. Rev.}\ {\bf D34}
                                       (1986), 722-736.
\bibitem{timeprecision}T.P. Heavner, S.R. Jefferts, E.A. Donley, J.H. Shirley \& T.E. Parker,
                       {\it Metrologia}\ {\bf 42} (2005), 411-422.
\bibitem{SalesiRecami1997}G. Salesi \& E. Recami, {\it Found. Phys. Lett.}\ {\bf 10} (1997), 533-546.
\end{thebibliography}
\end{document}